\documentclass[conference]{IEEEtran}
\IEEEoverridecommandlockouts
\usepackage{setspace}
\usepackage{cite}
\usepackage{amsmath,amssymb,amsfonts}
\usepackage{algorithmic}
\usepackage{graphicx}
\usepackage{textcomp}
\usepackage{xcolor}
\usepackage{array}
\usepackage{hyperref}
\usepackage{comment}
\usepackage{balance}
\usepackage{url}
\usepackage{acronym}
\input{Acronym}

\def\BibTeX{{\rm B\kern-.05em{\sc i\kern-.025em b}\kern-.08em
    T\kern-.1667em\lower.7ex\hbox{E}\kern-.125emX}}
\begin{document}
\setstretch{1}

\title{
NextG-GPT: Leveraging GenAI for Advancing Wireless Networks and Communication Research}

\author{\IEEEauthorblockN{Ahmad M. Nazar}
\IEEEauthorblockA{\textit{Elec. \& Comp. Eng. Dept.} \\
\textit{Iowa State University}\\
Ames, IA, USA \\
amnazar@iastate.edu}
\and
\IEEEauthorblockN{Mohamed Y. Selim}
\IEEEauthorblockA{\textit{Elec. \& Comp. Eng. Dept.} \\
\textit{Iowa State University}\\
Ames, IA, USA \\
myoussef@iastate.edu}
\and
\IEEEauthorblockN{Daji Qiao}
\IEEEauthorblockA{\textit{Elec. \& Comp. Eng. Dept.} \\
\textit{Iowa State University}\\
Ames, IA, USA \\
daji@iastate.edu}
\and
\IEEEauthorblockN{Hongwei Zhang}
\IEEEauthorblockA{\textit{Elec. \& Comp. Eng. Dept.} \\
\textit{Iowa State University}\\
Ames, IA, USA \\
hongwei@iastate.edu}
}

\maketitle

\begin{abstract}
Artificial intelligence (AI) and wireless networking advancements have created new opportunities to enhance network efficiency and performance. In this paper, we introduce Next-Generation GPT (NextG-GPT), an innovative framework that integrates \ac{RAG} and \acp{LLM} within the Wireless systems' domain. By leveraging state-of-the-art LLMs alongside a domain-specific knowledge base, NextG-GPT provides context-aware real-time support for researchers, optimizing wireless network operations. Through a comprehensive evaluation of LLMs—including Mistral-7B, Mixtral-8×7B, LLaMa3.1-8B, and LLaMa3.1-70B—we demonstrate significant improvements in answer relevance, contextual accuracy, and overall correctness. In particular, LLaMa3.1-70B achieves a correctness score of 86.2\% and an answer relevancy rating of 90.6\%. By incorporating diverse datasets such as ORAN-13K-Bench, TeleQnA, TSpec-LLM, and Spec5G, we improve NextG-GPT’s knowledge base, generating precise and contextually aligned responses. This work establishes a new benchmark in AI-driven support for next-generation wireless network research, paving the way for future innovations in intelligent communication systems.
\end{abstract}


\section{Introduction}
\label{introduction}
The rapid development of 5G and the anticipated emergence of 6G wireless networks have created a demand for more intelligent, autonomous, and adaptive network management solutions. Wireless communication research requires precise knowledge of telecom protocols, efficient network configurations, and real-time decision-making capabilities. However, traditional methods rely on manual configurations, static documentation, and iterative experimentation, making them time-consuming and resource-intensive. To address these challenges, AI and \acp{LLM} offer a promising solution for automated knowledge retrieval and AI-driven network analysis.

Despite their capabilities, general-purpose LLMs struggle with domain-specific tasks in wireless communications due to knowledge cutoffs, hallucinations, and an inability to handle domain-specific contexts \cite{ragSurvey, hallucinationSurvey}. \ac{RAG} mitigates these limitations by retrieving relevant, up-to-date information from structured knowledge bases, ensuring that responses are contextually accurate and technically grounded. 

To advance AI-driven wireless network research, we introduce Next-Generation GPT (NextG-GPT), a domain-specific \ac{RAG}-enhanced \ac{LLM} assistant designed for telecom applications, O-RAN research, and wireless experimentation while deploying it within a research testbed. NextG-GPT integrates multiple structured datasets to provide high-quality knowledge retrieval and intelligent response generation. This approach aligns with ongoing efforts to explore \acp{LLM} in the telecom domain and supports advancing the development of AI-driven wireless networks \cite{genai, transformer}.

The \textit{ARA Wireless Living Lab} (ARA) \cite{ara2}, based at Iowa State University, provides a large-scale testbed for advancing wireless communication technologies enabling real-world experimentation with next-generation network innovations. NextG-GPT integrates two datasets associated with ARA. The ARA documentation and APIs give users real-time access to technical documentation for configuring and troubleshooting network components. The Ericsson Base Station documentation facilitates configuring and managing base station operations within ARA using Moshell-based control systems \cite{ara2}. The TeleQnA dataset included is a benchmark designed to evaluate LLM understanding of telecommunications concepts \cite{teleqna}. ORAN-Bench-13K, a large-scale dataset for benchmarking LLMs in Open Radio Access Network (O-RAN) environments, is also included in NextG-GPT \cite{oran13k}. The TSpec-LLM and SPEC5G datasets provide extensive protocol and standards-related information that cover 3GPP telecom protocols and 5G network specifications, aiding in standards compliance and security analysis \cite{tspecllm, spec5g}.

These datasets allow NextG-GPT to support applications such as O-RAN benchmarking, telecom standards interpretation, network diagnostics, and AI-assisted experiment automation. NextG-GPT provides real-time assistance by utilizing \ac{RAG}-based retrieval and \ac{LLM}-driven generation for network optimization, configuration troubleshooting, and research decision-making. As such, this work evaluates the effectiveness of RAG-enhanced LLMs by systematically assessing their impact on response accuracy, contextual awareness, and practical usability within wireless and O-RAN environments.

The key contributions of this work are as follows:
\begin{itemize}
    \item \textbf{First \ac{RAG}-\ac{LLM} implementation in a wireless research testbed.} To our knowledge, NextG-GPT is the first deployment of an \ac{RAG}-based \ac{LLM} assistant within a next-generation wireless research environment, ARA, extending the role of AI in wireless experimentation.
    \item \textbf{Evaluation of \ac{LLM} performance in wireless contexts.} A comparative analysis of multiple \ac{LLM} architectures benchmarks their effectiveness in domain-specific tasks.
    \item \textbf{Integration of telecom-specific knowledge bases.} NextG-GPT leverages structured wireless network standards, O-RAN, and ARA research datasets to improve domain adaptation while reducing factual inconsistencies.
    \item \textbf{Advancement of AI-driven wireless research.} NextG-GPT accelerates innovation and enables automated network configuration, intelligent troubleshooting, and AI-powered knowledge retrieval.
\end{itemize}

This paper is organized as follows: Section \ref{Related-Work} summarizes related work. Section \ref{system-model} describes NextG-GPT’s system architecture. NextG-GPT evaluation methodologies and metrics are detailed in Section \ref{methodology}. Section \ref{eval_results} presents results and analysis. Section \ref{integrations} discusses use cases of NextG-GPT. Section \ref{challenges} discusses challenges with deploying \ac{RAG}-\acp{LLM} in wireless networks. Section \ref{Future_Work} discusses our research directions with NextG-GPT, and finally, Section \ref{conclusion} concludes our findings.

\section{Related Work}
\label{Related-Work}
This section summarizes previous work utilizing \ac{RAG}-\acp{LLM} techniques in domain-specific applications.

\subsection{RAG-LLM Assistants in Wireless System Development}
Recent research has explored the role of \acp{LLM} in optimizing wireless communication processes. The work in \cite{wirelessllm} introduces WirelessLLM, a framework designed to adapt \acp{LLM} for wireless intelligence by incorporating knowledge alignment, fusion, and evolution. The study examines key enabler technologies such as prompt engineering, \ac{RAG}, and domain-specific fine-tuning to enhance spectrum management, interference mitigation, and intelligent resource allocation.

Additionally, \acp{LLM} have been investigated for their potential to accelerate hardware design in wireless systems. The study in \cite{llm-fpga} explores the application of \acp{LLM} in FPGA-based hardware development for advanced wireless signal processing. Through this approach, \acp{LLM} improves development efficiency in complex wireless communication projects. 

\subsection{LLM Applications in Telecommunications}
Several studies have explored the role of \acp{LLM} in assisting telecommunications research and development. The work in \cite{tka} introduces the Telecom Knowledge Assistant (TKA), an \ac{RAG}-\ac{LLM}-based system designed to assist domain experts with technical queries related to 3GPP networking standards. While TKA offers valuable insights into \acp{LLM} for telecommunications, its focus is limited to standards-based documentation and does not extend to interactive research assistance or real-time experimentation support.

In \cite{llm-wireless}, the application of \acp{LLM} in wireless networks with prompt engineering techniques guide \acp{LLM} to generate accurate and context-aware responses, improve flexibility and resource efficiency. However, while prompt engineering enhances \ac{LLM} adaptability, it does not address the challenges associated with integrating domain-specific contexts.

\subsection{Innovation of NextG-GPT}
NextG-GPT is the first \ac{RAG}-\ac{LLM}-based assistant designed explicitly for real-time wireless experimentation, network optimization, and AI-driven research assistance. It uniquely integrates domain-specific telecom datasets, O-RAN benchmarks, and 5G/6G network documentation, providing context-aware insights beyond traditional information retrieval.

NextG-GPT moves beyond these approaches by integrating telecom-specific datasets, \ac{RAG}, and real-time experimental validation. Furthermore, its deployment within the \textit{ARA Wireless Living Lab} provides a real-world testbed for evaluating AI-driven wireless research, distinguishing it from previous domain-specific \ac{LLM} applications.


\section{Implementation of NextG-GPT}
\label{system-model}
\begin{figure}
    \centering
    \includegraphics[width=1.0\linewidth]{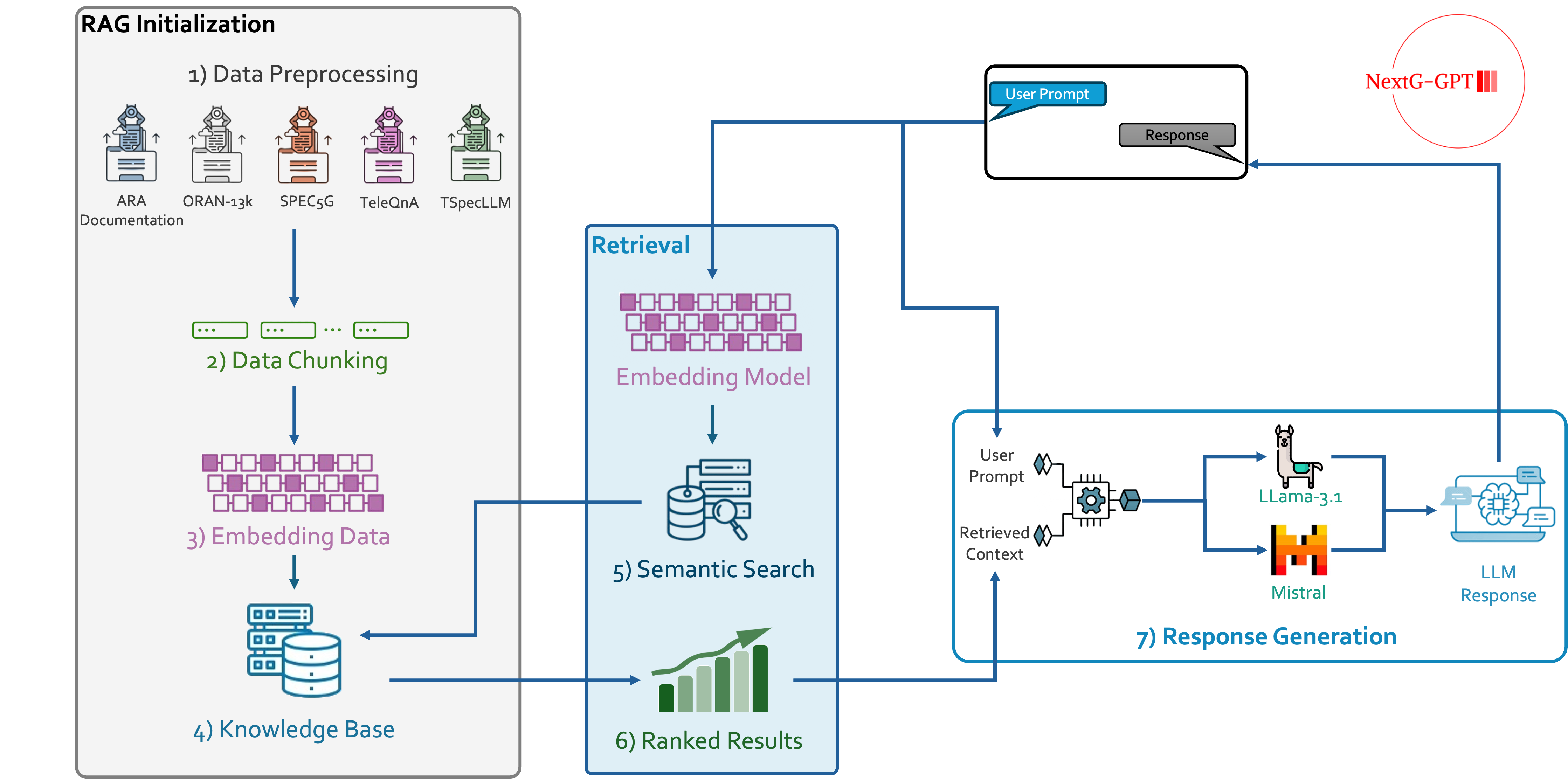}
    \caption{NextG-GPT workflow where Steps 1-4 involve RAG initialization; on prompting, Steps 5-6 show the semantic search and result ranking mechanism, and Step 7 shows the response generation where the user prompt and appropriate contexts are provided to the LLM to generate a response.}
    \label{fig:nextg-gpt-workflow}
\end{figure}

NextG-GPT leverages advanced \acp{LLM} with \ac{RAG} as its core framework. RAGs enable the integration of a knowledge base into the generative process, ensuring that the generated responses are contextually relevant and accurate.  
Figure \ref{fig:nextg-gpt-workflow} illustrates the steps in the NextG-GPT workflow, seamlessly integrating data extraction (Steps 1-2), text embedding (Step 3), knowledge base construction (Step 4), retrieval mechanisms (Step 5-6), and response generation (Step 7), which are discussed in detail in the following subsections.

\subsection{Datasets}
The knowledge base for NextG-GPT integrates multiple datasets to ensure precise configuration, optimization, and troubleshooting of wireless network components. The key datasets are as follows:

\begin{itemize}
    \item \textbf{ARA Documentation and APIs}: Comprehensive information on ARA, including its APIs, configuration guides, and technical specifications for Ericsson Base Stations, enabling researchers to configure and optimize network components effectively.
    \item \textbf{ORAN-Bench-13K}: A dataset comprising entries referenced from 116 O-RAN specification documents, providing detailed insights into Open Radio Access Network (O-RAN) standards.
    \item \textbf{TeleQnA}: A structured dataset curated from telecom-related knowledge extracted from technical standards, research papers, and industry documentation, designed to benchmark LLM's understanding of telecommunications concepts.
    \item \textbf{TSpec-LLM}: An open-source dataset covering all 3GPP documents from Release 8 to 19. It provides extensive information on cellular network protocols, standards, and regulatory frameworks, including compliance details.
    \item \textbf{SPEC5G}: A dataset containing natural language specifications extracted from 5G cellular network protocol documents, aiding in protocol analysis and 5G standard compliance.
\end{itemize}

\subsection{Embeddings}
\ac{RAG} effectiveness depends heavily on the quality and relevance of the retrieved passages, which is best compared using vectorized representations. NextG-GPT employs a pre-trained \ac{GTE} model, which supports a substantial context length of 8192 tokens, transforming datasets into high-dimensional vector representations. This encoding process maps textual information into a semantic space.
\ac{GTE} model was selected based on operational constraints and its ability to achieve optimal retrieval precision.

\subsection{LLMs}
\acp{LLM} serve as the generative component of NextG-GPT, processing retrieved contextual data alongside user prompts to generate relevant responses. These models excel in natural language processing tasks due to their advanced architecture, which predicts subsequent tokens in a manner that maintains coherence and relevance \cite{llmApps, decoder}.

To systematically evaluate how model size influences \ac{RAG}-enhanced performance, NextG-GPT utilizes several state-of-the-art \acp{LLM} ranging from 7 to 70 billion parameters. This selection enables an empirical comparison to determine whether larger models consistently deliver superior results when integrated with \ac{RAG}.

While smaller models provide computational efficiency, larger models are hypothesized to offer enhanced contextual reasoning and factual accuracy. Through empirical analysis, NextG-GPT assesses whether increasing model size leads to improved answer relevancy, correctness, and faithfulness, which are evaluation metrics described in Section \ref{eval_results} in domain-specific knowledge retrieval.

Furthermore, the choice of \acp{LLM} is guided by their open-source nature, allowing unrestricted customization and deployment.
Open-source models, including those from Mistral AI and Meta \cite{ mixtral, llama3.1}, provide flexibility and cost efficiency.


\subsection{NextG-GPT Workflow}
NextG-GPT’s performance fundamentally depends on a knowledge base integrating multiple wireless communications datasets, enabling real-time assistance for network experimentation, infrastructure configuration, and referencing telecommunications standards. Below are the steps in NextG-GPT's workflow as depicted by Fig. \ref{fig:nextg-gpt-workflow}.

\subsubsection{\textbf{Data Preprocessing}}
Datasets undergo a structured preprocessing pipeline to ensure consistency and usability. The text extraction process removes formatting irregularities, filters redundant content, and segments textual data into uniformly formatted sections optimized for retrieval-based tasks.

\subsubsection{\textbf{Data Chunking}}
Once the datasets are preprocessed, they are segmented into manageable and equal-sized chunks. This chunking process ensures the data is consistently integrated into the \ac{GTE} model.

Consider a preprocessed dataset of size ${L}$. This dataset is divided into smaller ${C}$-sized chunks. Each chunk overlaps with the next by ${C}_\text{o}$ characters to maintain continuity and prevent loss of meaning between chunks. 
The parameters, ${C} = 800$ and ${C}_\text{o} = \text{int}({C}/10) = 80$, were selected for chunking.


\subsubsection{\textbf{Embedding Data}}
After segmenting the preprocessed data into chunks, each chunk is transformed into a vectorized representation using a \ac{GTE} model. Text chunks are tokenized and then mapped to numerical vector representations. This final output is a fixed-size vector embedding

Ensuring consistent embedded vector size is achieved during the tokenization and embedding stages, where padding ensures uniform tokenized chunk lengths and consistent semantic representation. The final embedded vector with padding is matched to the \ac{GTE} model's maximum token size for simplicity. This uniformity allows \ac{LLM}s to handle redundancy, conflicts, and synergies effectively. 

\subsubsection{\textbf{Knowledge Base Creation}}
After each chunk is vectorized and embedded, they are appended to the knowledge base for retrieval. This integrated approach facilitates robust, conflict-free sensing and a unified understanding of the environment.

The \ac{FAISS} framework \cite{faiss} is used to store these vectorized representations. \ac{FAISS} provides similarity search frameworks that facilitate efficient indexing and retrieval of embeddings.

\subsubsection{\textbf{Semantic Search}}
On successful knowledge base initialization, users can now prompt NextG-GPT. Upon receiving a user prompt, NextG-GPT first tokenizes and embeds the prompt using the \ac{GTE} model. Once the prompt is vectorized, NextG-GPT conducts a semantic similarity search between the vectorized prompt and knowledge base entries. 
By calculating the semantic similarity, i.e., the cosine similarity between the embedded user prompt and the embedded data, NextG-GPT can fairly and quickly identify entries that closely match the user prompt and retrieve the most relevant contexts. 

FAISS optimizes vector similarity searches through hierarchical indexing and clustering, enabling efficient handling of large-scale embeddings. This indexing method ensures that top-ranked result retrieval operations remain low-latency even as data volume grows while ensuring high relevancy in retrieved contexts. 

\subsubsection{\textbf{Ranked Results}}
With retrieval, many vectorized embeddings may share similar characteristics, and as such, retrieved results are ranked and filtered to use the most relevant contexts. The ranking process employs top-$p$ percentile relevance filtering, where $p=95$ is consistently applied. In the context of \ac{RAG}, this step retains the top 95\% most relevant results based on their semantic similarity scores with the user prompt. FAISS optimizes ranking by sorting and scoring the retrieved context's scores through hierarchical indexing. These retrieved contexts are then decoded back into text by the \ac{GTE} model to be used with the user prompt as \ac{LLM} inputs.

\subsubsection{\textbf{Generated Response}}
At this step, NextG-GPT has extracted the relevant context and forwards it with the original textual user prompt to the \ac{LLM}. This information is processed by the \ac{LLM} to generate a contextually relevant response. As a response generation enhancement, NextG-GPT utilizes top-$p$ sampling, where $p = 95$ is selected as a parameter for consistency, to balance accuracy and diversity in responses while choosing the most appropriate response and maintaining contextual relevance. Top-$p$ sampling chooses from the smallest possible set of words whose cumulative probability exceeds the probability $p$. These filtered samples are then used to generate a contextually accurate and appropriate response.

\section{LLM Evaluation Methodology}
\label{methodology}
Establishing clear and measurable criteria to evaluate the efficacy of the employed LLMs is essential for NextG-GPT. LLMs are generally assessed on \ac{GLUE} and \ac{MMLU} benchmarks. However, these assessments can be too general. For a fair evaluation of domain-specific approaches, specific metrics are utilized to assess the performance of the \acp{LLM}, including answer relevancy, context recall, correctness scores, and faithfulness, as found in the RAGAS Evaluator \cite{ragas}. Each metric provides valuable insights into NextG-GPT's capabilities through different LLMs to comprehensively analyze their practicalities. Below, we detail the methodologies for computing these metrics.

\subsubsection{\textbf{Answer Relevancy (AR)}}
Answer relevancy evaluates how well the generated response aligns with its retrieved-context and ground truth. This metric is crucial as misinterpretations can lead to inefficiencies or operational errors in deployment. AR is measured by the average cosine similarity between the generated response and its corresponding ground truth, defined as:
\begin{equation}
\text{AR}=\frac{1}{N} \sum_{i=1}^N \cos\left(\vec{E}_{p_i}, \vec{E}_{t_i}\right) =\frac{1}{N} \sum_{i=1}^N \frac{\vec{E}_{p_i} \cdot \vec{E}_{t_i}}{\left\|\vec{E}_{p_i}\right\|_2\left\|\vec{E}_{t_i}\right\|_2},
\end{equation}
where $\vec{E}_{p_i}$ and $\vec{E}_{t_i}$ represent the vectorized embeddings of the $i$-th generated response and its ground truth, respectively.

\subsubsection{\textbf{Context Recall}}
Context recall measures the alignment between the retrieved context and the ground truth, ensuring that retrieved information is relevant and contributes meaningfully to response generation. It is computed by normalizing the overlap between retrieved-context sentences and those in the ground truth.

\subsubsection{\textbf{Correctness (AC)}}
Correctness assesses response accuracy by comparing the generated answer with the ground truth. It is computed as a weighted sum of semantic similarity and factual correctness.

Semantic similarity measures how closely the generated response resembles the ground truth in meaning. The embeddings of the ground truth ($\vec{E}_{t_i}$) and the generated response ($\vec{E}_{a_i}$) are computed, and their cosine similarity is used to quantify their semantic closeness.

Factual correctness determines the factual overlap between the generated and ground truth responses:
\begin{equation}
F = \frac{|\text{TP}|}{(|\text{TP}| + 0.5 \times (|\text{FP}| + |\text{FN}|))},
\end{equation}
where TP represents true positives (statements present in both responses), FP denotes false positives (statements in the generated response but not in the ground truth), and FN refers to false negatives (statements in the ground truth but missing in the generated response).

Correctness is then defined as:
\begin{equation}
\text{AC}=\omega \cos\left(\vec{E}_{a_i}, \vec{E}_{t_i}\right) + (1-\omega) F,
\end{equation}
where $\omega$ is a weight factor balancing semantic similarity and factual correctness. For our evaluation, $\omega = 0.25$.

\subsubsection{\textbf{Faithfulness (AF)}}
Faithfulness evaluates factual consistency, ensuring all generated claims are logically inferred from the retrieved context. It is defined as:
\begin{equation}
    \text{AF}=\frac{|N_{Gc}|}{|N_C|},
\end{equation}
where $N_{Gc}$ represents the number of claims in the response supported by the given context, and $N_C$ is the total number of claims in the response.

\section{LLM Evaluation Results}
\label{eval_results}
Several metrics were analyzed, and several models were employed to produce diverse results to evaluate Next-GPT. NextG-GPT was evaluated using the RAGAS evaluator on Mistral-7b (M7b), Mixtral-8x7b (M47b), LLaMa3.1-8b (L8b), and LLaMa3.1-70b (L70b) \cite{mistral, mixtral, llama3.1}. The model sizes are measured in billions of parameters, 7, 47, 8, and 70 billion, respectively.
We utilize Vanilla LLaMa3.1-70b (V-L70b) and Mixtal-8x7b (V-M47b) to compare off-the-shelf \acp{LLM} and \ac{RAG}-\acp{LLM}. These models were selected for their competitive performance and ability to generate responses. We utilize stella\_en\_400M\_v5 from NovaSearch as the \ac{GTE} model \cite{stella}.

Our evaluation metrics provide insight into NextG-GPT's capabilities based on the LLMs used. Four test sets containing questions related to each dataset, their ground truths answers, and their contexts were created to evaluate NextG-GPT's performance. Each test set consisted of $N=30$ question and answer pairs. The scores herein are ordered based on their scores in the respective datasets: ARA Wireless Combined Documentation, Spec5G, ORAN-Bench-13K, TeleQnA, and TSpec-LLM. It is worth noting that vanilla LLMs do not include results for answer relevancy, faithfulness, and context recall as they do not utilize a knowledge base and thus cannot be evaluated among those metrics. 

\begin{figure*}
    \centering
    \includegraphics[width=1.0\linewidth]{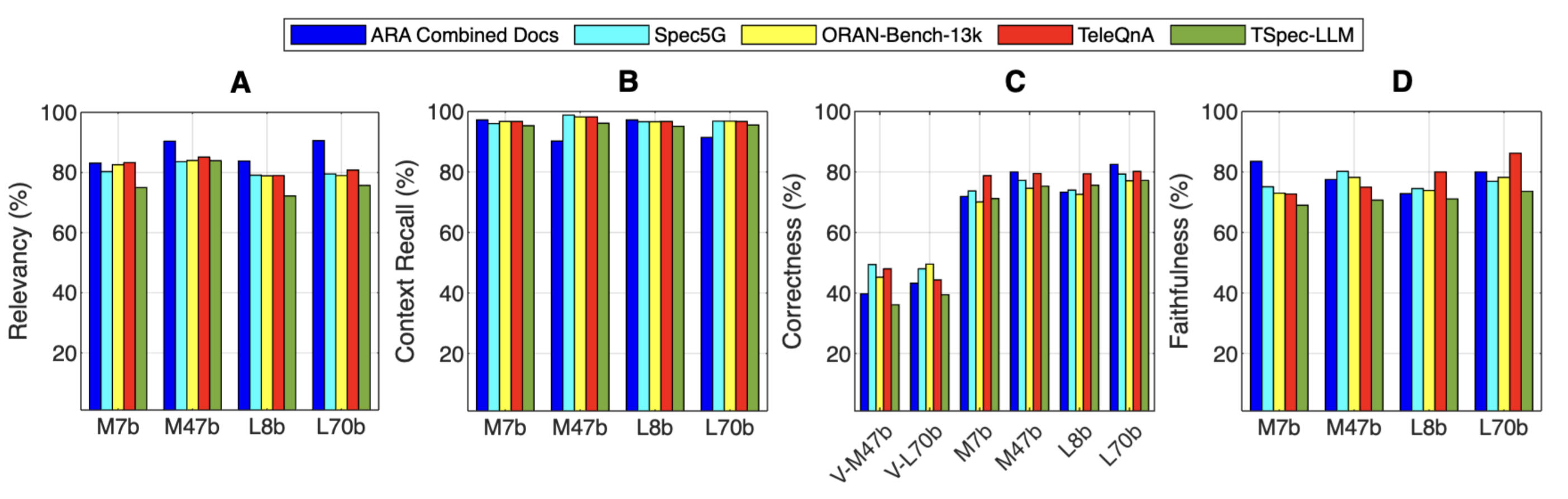}
    \caption{Evaluation Metrics of NextG-GPT where A) shows answer relevancy scores, B) shows context recall scores, C) includes vanilla LLM and RAG-LLM answer correctness scores, and D) shows answer faithfulness.}
    \label{results}
    \vspace{-10pt}
\end{figure*}

\subsection{Answer Relevancy}
NextG-GPT's answer relevancy scores, shown in Fig. \ref{results}A, demonstrate that Mistral-7B scores 83.1\%, 80.3\%, 82.6\%, 83.3\%, and 75.0\%, demonstrating strong contextual relevance across datasets. LLaMa3.1-8B achieves 83.8\%, 79.1\%, 78.9\%, 79.0\%, and 72.2\%, showing consistent but slightly lower relevance. Mixtral-8x7B surpasses the smaller models with higher scores of 90.4\%, 83.6\%, 84.0\%, 85.1\%, and 83.9\%, indicating superior contextual understanding. LLaMa3.1-70B attains 90.6\%, 79.5\%, 79.0\%, 80.8\%, and 75.7\%, showcasing superior strength in providing relevant responses.

These results suggest that larger models tend to achieve higher answer relevancy, indicating their ability to understand and retain domain-specific context more effectively. The performance gap between smaller and larger models highlights the role of parameter scaling and dataset integration in refining contextual accuracy. Furthermore, RAG-based retrieval ensures that generated responses remain aligned with technical queries, reducing hallucinations and improve research usability in wireless communications.



\subsection{Context Recall}
NextG-GPT's context recall results, depicted in Fig. \ref{results}B, show that Mistral-7B achieves 97.2\%, 96.0\%, 96.7\%, 96.7\%, and 95.3\%, demonstrating consistent retrieval accuracy. Mixtral-8x7B scores slightly higher in some datasets, particularly Spec5G and ORAN-Bench-13K, with values of 90.2\%, 98.8\%, 98.2\%, 98.2\%, and 96.1\%. LLaMa3.1-8B maintains strong recall capabilities at 97.2\%, 96.6\%, 96.6\%, 96.7\%, and 95.1\%. LLaMa3.1-70B follows closely with 91.4\%, 96.8\%, 96.8\%, 96.7\%, and 95.5\%.

The high context recall scores suggest that NextG-GPT effectively retrieves relevant contexts, reinforcing the importance of structured knowledge bases in improving response accuracy. The slight variations in recall scores across models indicate that while larger models improve answer relevancy, retrieval efficiency remains high even in smaller architectures. 



\subsection{Correctness Scores}
NextG-GPT's correctness scores, shown in Fig. \ref{results}C, demonstrate that Mistral-7B scores 71.9\%, 73.7\%, 70.1\%, 78.8\%, and 71.2\%, showcasing reliable but variable accuracy. Mixtral-8x7B improves upon this with 80.0\%, 77.2\%, 74.6\%, 79.5\%, and 75.3\%, indicating greater consistency. LLaMa3.1-8B achieves 73.3\%, 74.0\%, 72.6\%, 79.4\%, and 75.6\%, performing similarly to Mixtral-8x7B. LLaMa3.1-70B achieves the highest scores at 82.5\%, 79.3\%, 77.1\%, 80.2\%, and 77.2\%, demonstrating superior accuracy.

In contrast, the vanilla models show significantly lower correctness scores. Vanilla Mixtral-8x7B achieves 39.7\%, 49.4\%, 45.2\%, 48.0\%, and 36.1\%, while Vanilla LLaMa scores 43.2\%, 48.0\%, 49.5\%, 44.3\%, and 39.4\%. 

These findings emphasize that \ac{RAG} is essential in reducing misinformation and improving factual accuracy. The stark contrast between \ac{RAG}-based and vanilla models demonstrates that access to structured knowledge bases significantly enhances model reliability. Additionally, the improvement in correctness across larger models suggests that parameter scaling plays a role in improving factual accuracy, but the biggest gains come from contextual knowledge retrieval rather than intrinsic model training alone.



\subsection{Faithfulness}
NextG-GPT's faithfulness scores, detailed in Fig. \ref{results}D, show that Mistral-7B achieves 83.5\%, 75.1\%, 73.0\%, 72.7\%, and 69.0\%, demonstrating strong contextual fidelity. Mixtral-8x7B attains 77.5\%, 80.2\%, 78.2\%, 75.0\%, and 70.7\%, indicating improved consistency. LLaMa3.1-8B scores 72.8\%, 74.5\%, 73.9\%, 80.0\%, and 71.1\%, with particularly strong performance in \textit{TeleQnA}. LLaMa3.1-70B achieves the highest faithfulness scores at 80.0\%, 76.9\%, 78.2\%, 86.2\%, and 73.6\%, highlighting superior accuracy in reflecting retrieved information.

Faithfulness is critical in ensuring that \acp{LLM} do not misinterpret or distort retrieved content, and these results indicate that NextG-GPT maintains strong alignment with its retrieved knowledge sources. The high faithfulness scores suggest that \ac{RAG}-based models effectively minimize fabrication and ensure that AI-generated responses remain grounded in telecom-specific datasets. Additionally, the increased faithfulness in larger models suggests that they process retrieved contexts with greater coherence.



\subsection{Generated Response}
\begin{figure*}
    \centering
    \includegraphics[width = 1.0\linewidth]{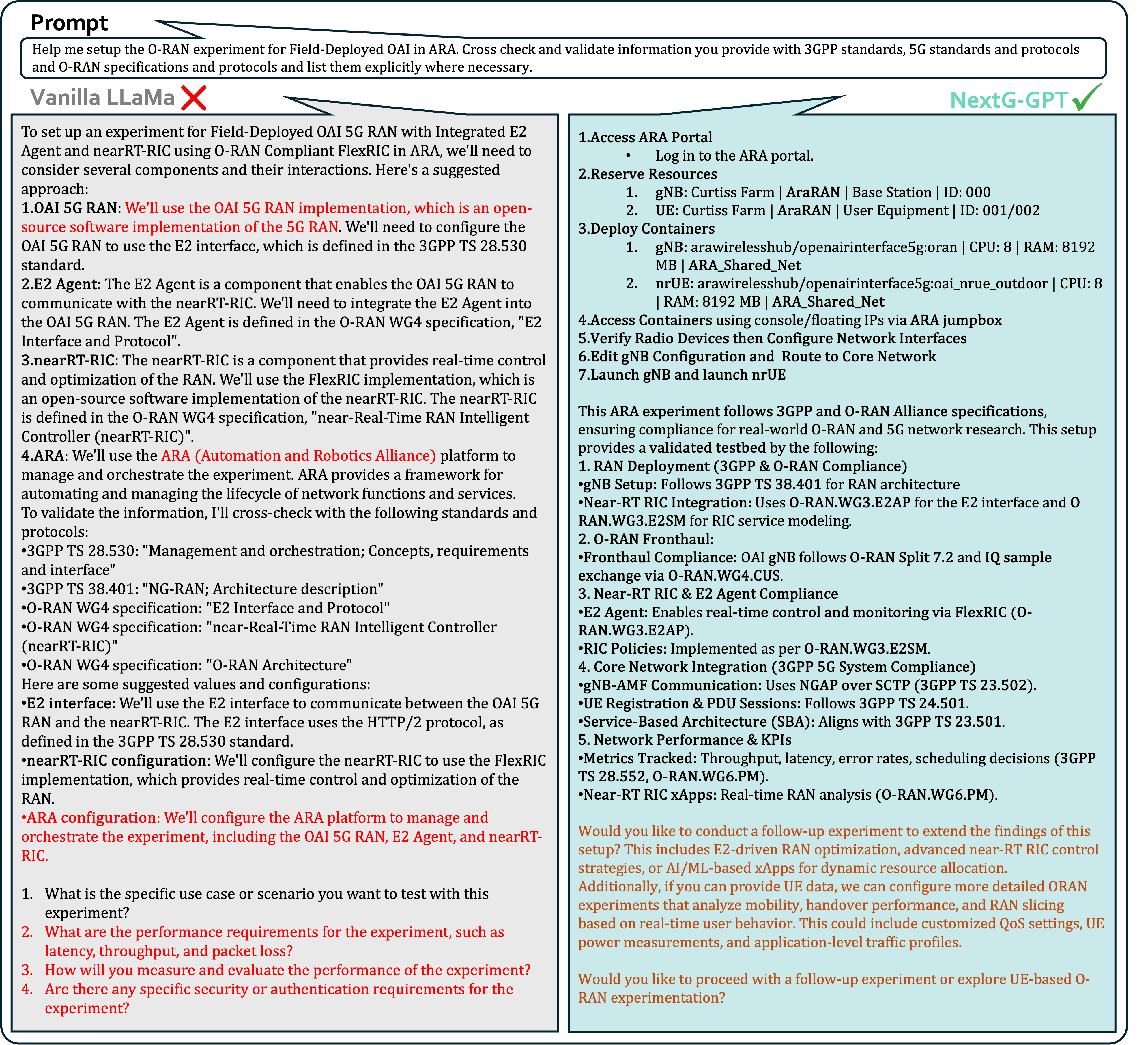}
    \caption{Comparison of Vanilla LLaMa and NextG-GPT responses to an ARA-specific O-RAN experiment setup query. NextG-GPT provides accurate, structured instructions with validated 3GPP and O-RAN references, while Vanilla LLaMa gives a generic and partially incorrect response, as shown in the red text.}
    \label{fig:example_results}

\end{figure*}

\ac{RAG} effectiveness is illustrated in Fig. \ref{fig:example_results}, where an ARA-specific prompt submitted to Vanilla LLaMa and NextG-GPT clearly distinguishes performance, as the red-colored text displays inaccuracies. Vanilla LLaMa fails to provide a usable response, while NextG-GPT accurately references the datasets used in NextG-GPT and outlines ARA portal access with deployment steps. This response demonstrates NextG-GPT’s superior retrieval capabilities, domain adaptation, and real-world usability. 

When prompted with an ARA-specific experiment setup request, Vanilla LLaMa generated a generic and partially incorrect response, failing to recognize ARA Wireless Living Lab and instead referring to the Automation and Robotics Alliance. This misinterpretation fundamentally compromised the relevance of its response, as it did not provide helpful information regarding the actual ARA testbed environment.

Beyond incorrect context, Vanilla LLaMa’s response was broad and lacked actionable instructions. Instead of providing a structured experimental setup, it described OAI 5G RAN, nearRT-RIC, and E2 Agent in generic terms without detailing how these components integrate within ARA. Furthermore, while it referenced 3GPP and O-RAN standards, it did so without applying them to the experiment setup, making its response detached from practical implementation. Additionally, its response included open-ended follow-up questions, shifting the burden onto the user rather than providing a straightforward and validated experimental procedure.

In contrast, NextG-GPT strictly adhered to the prompt’s requirements, demonstrating a deep understanding of ARA and the necessary O-RAN experiment setup. It delivered an explicit, structured workflow, including resource reservations, container deployments, and network configurations. Unlike Vanilla LLaMa, which merely mentioned standards, NextG-GPT directly integrated and cross-checked relevant 3GPP and O-RAN specifications into the response. 
For example, it correctly referenced 3GPP TS 38.401 for RAN architecture, O-RAN.WG3.E2AP for E2 interfaces and 3GPP TS 23.501 for service-based architecture compliance, ensuring that the provided steps align with real-world implementation guidelines.

NextG-GPT demonstrated superior \ac{RAG} capabilities by referencing the domain-specific datasets to synthesize precise technical insights while validating its recommendations with authoritative telecom standards. Unlike Vanilla LLaMa, which failed to provide direct setup instructions, NextG-GPT correctly referenced specifications and detailed the ARA portal access process, demonstrating its retrieval accuracy and domain adaptation.

\subsection{Discussions}
Our analysis reveals significant variations in evaluation metrics among the selected \acp{LLM}. These differences provide valuable insights into optimizing retrieval mechanisms and refining workflow efficiency in wireless network research.

One key observation is that Mixtral-8x7B performs comparably to LLaMa3.1-70B, demonstrating that model efficiency is not solely dependent on parameter size. This finding suggests that architectural innovations in smaller models can achieve performance levels similar to significantly larger models while reducing computational overhead.

Models with higher scores in the metrics demonstrate substantial improvements in delivering precise and contextually relevant responses. This precision enhances workflow efficiency by reducing the time required to retrieve accurate information, streamlining experiment setup, and minimizing errors. Additionally, the high answer relevancy observed in these models ensures smoother, more interactive dialogue exchanges, making NextG-GPT a more effective assistant in wireless network research.

Beyond optimizing workflows, NextG-GPT significantly enhances research retrieval and decision support by integrating diverse datasets tailored for wireless communications and networking research. The combination of \ac{RAG}-\acp{LLM} ensures that NextG-GPT effectively extracts, synthesizes, and delivers precise, up-to-date technical insights.


\section{LLM-Based Unified Solution for ARA Wireless Network Integrations}
\label{integrations}
NextG-GPT introduces a transformative AI-driven approach to wireless network research by integrating \ac{RAG}-\acp{LLM}. This section details use cases applicable to NextG-GPT.

\subsection{AI-Enhanced Experimentation and Network Optimization}
One of the primary applications of NextG-GPT is its role as an AI-enhanced experimentation assistant. Users expand its capabilities to dynamically design, configure, and optimize wireless network experiments. By analyzing real-time and historical network data, NextG-GPT recommends network parameter tuning, including frequency allocations, power levels, and protocol optimizations. The system also validates experimental setup by detecting configuration inconsistencies and ensuring alignment with research objectives. Furthermore, it enables adaptive experiment refinement by suggesting modifications based on observed performance metrics, reducing manual intervention and iterative testing.

\subsection{Intelligent Debugging and Fault Diagnosis}
NextG-GPT functions as an intelligent debugging and fault diagnosis assistant, addressing the complexities of troubleshooting large-scale wireless systems. Processing telemetry data, configuration logs, and performance metrics identifies the root causes of failures, such as radio frequency interference, misconfigurations, or protocol mismatches. Through interactive debugging, researchers can query the system in natural language to obtain step-by-step troubleshooting guidance, mitigating the need for exhaustive manual searches through documentation. Additionally, NextG-GPT supports predictive maintenance by identifying degradation patterns and recommending proactive interventions for network hardware, minimizing downtime and performance degradation.



\subsection{Autonomous Experimentation and AI-Driven Optimization}
NextG-GPT further extends its functionality into autonomous experimentation and optimization, leveraging reinforcement learning-based approaches to refine experimental configurations dynamically. By analyzing the impact of various parameters on performance metrics, it autonomously suggests modifications to optimize network throughput, latency, and reliability. Additionally, it facilitates automated hypothesis testing by simulating different configurations before real-world deployment, expediting the research cycle. 


\section{Implementation Challenges}
\label{challenges}
Despite its promising capabilities, NextG-GPT faces several challenges that must be addressed to ensure optimal performance and reliability in wireless network research. These challenges primarily concern computational efficiency, scalability, data quality, response accuracy, and deployment constraints.

\subsection{Memory Management and Computational Efficiency}
Deploying NextG-GPT requires substantial computational resources, particularly for large-scale \acp{LLM} exceeding 10 billion parameters. The reliance on extensive GPU memory and high inference costs can lead to bottlenecks, affecting real-time responsiveness and accessibility for researchers. Additionally, integrating \ac{RAG} introduces further computational overhead, as it involves indexing large knowledge bases, performing similarity searches, and ranking retrieved documents before generation. In resource-constrained environments, inefficient memory allocation can lead to latency issues, system instability, or failures in handling concurrent queries. Optimizing memory usage through quantization techniques, efficient batching strategies, and distributed inference architectures is essential to ensure the system remains responsive and scalable.

\subsection{Scalability and Adaptability}
As the volume and complexity of research queries grow, ensuring NextG-GPT’s scalability and adaptability remains a significant challenge. The system must efficiently handle increasing diverse queries while maintaining accuracy and responsiveness. One major scalability concern is retrieval efficiency, as the size of the knowledge base expands over time. Although \ac{FAISS} employs hierarchical indexing to improve search performance, retrieval latency may still increase due to the high-dimensional nature of vector searches. Additionally, as telecommunications research evolves, NextG-GPT must continuously adapt to new standards, protocols, and datasets to remain relevant. 

\subsection{Data Quality, Relevance, and Knowledge Base Maintenance}
NextG-GPT’s reliability directly depends on the quality and accuracy of the underlying knowledge base. Inconsistencies or inaccuracies in retrieved documents can lead to incorrect or misleading responses. One challenge lies in maintaining dataset integrity as new standards are introduced. The knowledge base must be continuously updated with new research findings, regulatory guidelines, and technical specifications to ensure relevance. Additionally, managing conflicting information from multiple sources presents another difficulty, as different entities often revise and reinterpret wireless communication standards. Implementing systematic dataset validation pipelines, automated knowledge ingestion mechanisms, and contradiction detection models is necessary to enhance NextG-GPT’s response reliability and credibility.

\subsection{Mitigating Hallucination and Response Uncertainty}
Hallucination remains a persistent issue in \ac{LLM}-based systems, where the model generates factually incorrect but seemingly plausible responses \cite{hallucinationSurvey}. In highly technical domains such as wireless networking, hallucinations can introduce errors in research guidance, mislead experiment configurations, or cause inaccuracies in telecom standard interpretations. Even with \ac{RAG} integration, hallucinations can still occur if the retrieved knowledge base entries are insufficient or incomplete, forcing the model to infer missing information. NextG-GPT can mitigate hallucination risks by implementing confidence-aware response filtering, where the model flags uncertain responses based on retrieval coverage. Additionally, leveraging uncertainty estimation techniques and cross-referencing generated outputs against external authoritative sources can enhance response trustworthiness. Establishing a mechanism for researchers to verify and flag incorrect outputs will also contribute to refining the system over time.

\section{Future Work}
\label{Future_Work}
As NextG-GPT continues to evolve, future enhancements focus on expanding its capabilities beyond text-based retrieval to enable more adaptive and autonomous network operations. The following subsections outline two key areas of innovation: multi-modal data integration for real-time situational awareness, adaptive RAN optimization, and autonomous wireless experimentation for self-optimizing networks.

\subsection{Multi-Modal Data Integration and Adaptive RAN Optimization}
While NextG-GPT has demonstrated strong performance in domain-specific knowledge retrieval, its reliance on textual inputs limits its ability to interpret dynamic wireless environments and optimize RAN behavior in real-time. A key direction for future development is the integration of multi-modal data sources such as real-time network telemetry, spectrum scans, LiDAR, GPS, and imaging data to enhance contextual awareness. By incorporating these data streams, NextG-GPT can move beyond passive retrieval and actively sense, analyze, and optimize network performance.

One critical application of multi-modal awareness is E2-driven RAN optimization, where NextG-GPT facilitates real-time control loops between the near-RT RIC and RAN nodes. NextG-GPT can assist in dynamic resource allocation, power control, and beamforming adjustments based on real-time network conditions. For example, spectrum scans and interference maps can guide adaptive power management and handover strategies, ensuring that RAN resources are allocated efficiently. 

Additionally, UE-based experimentation could leverage multi-modal sensing to analyze mobility, handover performance, and RAN slicing based on real-time user behavior. If researchers provide UE data, NextG-GPT could assist in configuring customized QoS settings, tracking power measurements, and monitoring application-level traffic profiles, as depicted by the orange text in Fig. \ref{fig:example_results}. By correlating network performance with user movement and environmental factors, it could identify optimal mobility strategies, detect anomalies in connectivity, and fine-tune RAN parameters to improve the quality of experience.

Integrating these capabilities within the ARA allows NextG-GPT to continuously monitor network performance, detect interference sources, and visualize real-time spectrum utilization. Extracting insights from environmental data could explain network degradations, predict performance bottlenecks, and recommend targeted optimizations. These insights bridge the gap between raw data and actionable intelligence, creating a more transparent, interpretable AI system that enhances research efficiency and network adaptability.

\subsection{Autonomous Wireless Experimentation and Self-Optimizing Networks}
Beyond improving situational awareness, an innovative extension of NextG-GPT is the development of an autonomous wireless experimentation and self-optimization framework, where the model assists researchers and actively designs, executes, and optimizes wireless experiments in real-time. 

Researchers manually configure experiments, adjust testbed parameters, and analyze results. NextG-GPT could be extended to generate experimental hypotheses autonomously, suggest network configurations, and execute real-time tests in controlled environments. Leveraging its \ac{RAG}, it could dynamically adjust experimental variables, compare outcomes with theoretical predictions, and iteratively refine network parameters for optimal performance.

For example, in a self-optimizing 5G/6G testbed, NextG-GPT could autonomously modify power levels, beamforming strategies, or spectrum allocation based on live performance metrics. By continuously learning from its adjustments and refining configurations through reinforcement learning, it could create an AI-driven closed-loop optimization system, minimizing human intervention while maximizing network efficiency.

Moreover, this capability could extend to automated protocol validation and anomaly detection. NextG-GPT could generate test cases for new communication protocols, execute simulations, and verify compliance with standards such as O-RAN or 3GPP. If deviations or security vulnerabilities are detected, it could propose countermeasures, acting as an AI-driven regulatory compliance and security assurance assistant.

The long-term vision for this capability is an AI-powered self-orchestrating network that autonomously manages itself in real-time, learns from past experiments, adapts to new conditions, and fine-tunes its performance dynamically. This paradigm shift would transform NextG-GPT from a static knowledge assistant into a fully autonomous research collaborator capable of designing, executing, and improving next-generation wireless networks with minimal human input.

\section{Conclusion}
\label{conclusion}

Integrating NextG-GPT into the ARA Wireless Living Lab is crucial in applying AI-driven tools for Next-G wireless networks. With RAG-LLMs, NextG-GPT delivers precise, contextually relevant, and up-to-date information, enhancing the research capabilities within ARA and communications research. Our evaluation underscores the importance of answer relevancy, context recall, answer correctness, and faithfulnuss for optimal performance. The diverse knowledge base enables NextG-GPT to support researchers in their experiments effectively. This research sets the foundation for future innovations where AI-driven systems like NextG-GPT could become integral in managing increasingly complex and dynamic wireless environments, including the advent of 6G networks and smart, connected infrastructure.

\balance

\section*{Acknowledgements}
This work is supported in part by the NSF awards 2130889, 2112606, 2212573, 2229654, and 2232461. The authors thank the members of the ARA team for contributing and supporting NextG-GPT.

\bibliographystyle{Bibliography/IEEEtran}
\bibliography{Bibliography/bibliography}

\end{document}